\documentstyle[12pt, epsfig] {article}
\def\Journal#1#2#3#4{{#1} {\bf #2}, (#4), #3}

\def\NPB{{\em Nucl. Phys.} B}
\def\PLB{{\em Phys. Lett.}  B}
\def\PRL{\em Phys. Rev. Lett.}
\def\PRD{{\em Phys. Rev.} D}

\def\be{\begin{equation}}
\def\ee{\end{equation}}
\def\bea{\begin{eqnarray}}
\def\eea{\end{eqnarray}}

\makeatletter
\def\gsim{\compoundrel>\over\sim}
\def\lsim{\compoundrel<\over\sim}
\def\compoundrel#1\over#2{\mathpalette\compoundreL{{#1}\over{#2}}}
\def\compoundreL#1#2{\compoundREL#1#2}
\def\compoundREL#1#2\over#3{\mathrel
         {\vcenter{\hbox{$\m@th\buildrel{#1#2}\over{#1#3}$}}}}
\makeatother

\makeatletter

\newcommand{\lm}[1]{\mbox{$\lambda_{#1}$}}
\newcommand{\m}[1]{\mbox{$m_{#1}^{2}$}}
\newcommand{\p}[1]{\mbox{$\phi_{#1}$}}
\renewcommand{\H}[1]{\mbox{$H_{#1}$}}
\newcommand{\Hd}[1]{\mbox{$H_{#1}^{\dag}$}}

\begin{document}

\begin{flushright}
GUTPA/01/05/01
\end{flushright}
\vskip .1in

\begin{center}
{\Large {\bf Light Higgs Boson in the Spontaneously CP Violating NMSSM}}

\vspace{35pt}

{\bf A.T. Davies, C.D. Froggatt and A. Usai}

\vspace{6pt}

{\em Department of Physics and Astronomy\\[0pt]
Glasgow University, Glasgow G12 8QQ, Scotland\\[0pt]
}

\bigskip

{\large {\bf Abstract}} 
\end{center}
We consider spontaneous CP violation in the Next to Minimal Supersymmetric 
Standard Model, without the usual $Z_3$ discrete symmetry. CP violation can
occur at tree level, raising a potential conflict with the experimental 
bounds on the electric dipole moments of the electron and neutron. One 
escape from this is to demand that the CP violating angles are small, but 
we find that this entails a light neutral Higgs particle. 
This is almost pseudoscalar, can have a high singlet content, 
and will be hard to detect experimentally.

\thispagestyle{empty} \newpage

\section {Introduction} \label{Intro}

Although the observed lack of CP symmetry is readily accommodated in the 
Standard Model as a generic feature of the 3x3 CKM matrix,  there are 
other ways in which CP non-conservation can be introduced. Additional 
sources of CP violation are required to make electroweak baryogenesis 
viable, and this could be provided by the Higgs sector.

CP can be violated in the Higgs sector either explicitly, through complex 
coupling constants in the Lagrangian, or spontaneously, when, although the 
couplings are real, fields acquire complex vacuum expectation values (vevs). 
It 
should be acknowledged that spontaneous breaking of CP gives rise to domain 
walls which cause a cosmological problem, particularly if they are formed 
relatively late at the electroweak scale. Some suggestions for circumventing 
this problem have been made 
\cite{Krauss}.

At tree level neither type of CP violation occurs in the Standard Model or 
the Minimal Supersymmetric Standard Model (MSSM). 
In the MSSM  two or more complex 
phases can be explicitly introduced in the soft Susy-breaking potential, and 
at one loop these may give rise to  embarrassingly large  electric dipole 
moments for the neutron and electron. It has been shown that cancellations 
between contributing diagrams can reduce these dipole moments to within 
experimental bounds in a significant region of the parameter 
space~\cite{IbNath,Kane}.
However, a recent analysis~\cite{Abel} incorporating new data 
on mercury atoms  
suggests that small phases are still required.  CP can also be violated 
spontaneously due to radiative 
corrections, but because CP is conserved at tree level the CP phases on the 
Higgs fields are small, and a light almost pseudoscalar particle is 
predicted~\cite{Maekawa,Pom1}, in accordance 
with the Georgi-Pais theorem~\cite{Georgi}.  This boson is ruled out by LEP 
experiments~\cite{Gay}.

In Susy models the next simplest case is the Next to Minimal Standard Model 
(NMSSM), in which a gauge singlet scalar field is present in addition to the 
standard two doublets. In the most commonly discussed version of the NMSSM, 
which 
has a discrete $Z_3$ symmetry, Spontaneous CP Violation (SCPV) 
does not occur at tree level~
\cite{Romao}. Breaking can occur by radiative corrections but tends to 
predict a light scalar as in the MSSM~\cite{Babu}.

In a previous paper we considered ~\cite{DavJer} a 
more general NMSSM, with no  discrete $Z_3$ symmetry, and 
found that this allows spontaneous CP violation at tree level, without 
concomitant light neutral bosons. With such a spectrum the CP violating phases 
are large, and generate large contributions to electric dipole moments. 
Cancellations can be arranged so as to be consistent with experiment, 
but require fine tuning of some soft 
Susy-breaking terms in the potential.

A more interesting possibility is that CP is spontaneously broken only 
weakly, in the sense that the phases on the vevs are small.
For phases $\lsim 0.01$ radians the predicted electric dipole moments 
are suppressed sufficiently even without cancellation~\cite{Abel}. 
In this case 
we predicted a light almost purely pseudoscalar boson~\cite{DavCop}, 
a result recently confirmed in ~\cite{Branco}. 
Its  existence does not follow from the Georgi-Pais theorem but 
may be understood by a similar 
argument. We sample the space of unknown coupling parameters and consider 
the detectability of such a particle. In most cases this particle 
contains a high proportion of singlet field, which does not couple to 
gauge bosons or quarks, so it may well escape experimental detection.

After introducing our model we give a qualitative argument for the 
existence of a light pseudoscalar in the case of weak SCPV. 
Imposing small CP phases but otherwise 
relatively little theoretical 
bias, we randomly sample the large space of unknown parameters and find 
typical masses and couplings of the Higgs scalars. We then consider the 
experimental constraints.

\section{NMSSM} \label{NMSSM}

Our model is based on the superpotential
\begin {eqnarray}
W =  \lambda NH_1H_2 -\frac{k}{3} N^3 - r N + \mu H_1H_2 + W_{Fermion}
\end{eqnarray}
where $H_1$ and $H_2$ are the doublets of the MSSM and $N$ is a singlet.
A possible quadratic $N^2$ term has been removed by a field shift~
\cite{Gunion}.
Traditionally the NMSSM has been studied as a possible solution to the 
$\mu$-problem of the MSSM.  If the N field acquires a vev $x$ of 
the same scale as those of  $H_1$ and $H_2$, then  $\lambda x$ 
in the term $\lambda N H_{1}\,H_{2}$ provides a 
$\mu$ of the electroweak scale rather than the GUT scale.  We adopt a 
different viewpoint and regard the NMSSM as just a phenomenological 
generalisation of the MSSM.
We do not impose the usual discrete $Z_3$ symmetry in which ($H_1,H_2,N$) are 
rephased by $\exp(i2\pi/3)$, which would require $\mu = r = 0$.  

At the electroweak scale the effective potential is~\cite{Higgshunter}
\begin{eqnarray}
\label{eq.vs0} 
V_{0} & = & \frac{1}{2}\lm{1}(\Hd{1}\H{1})^2 + \frac{1}{2}\lm{2}
(\Hd{2}\H{2})^2   \nonumber  \\
  &  & + (\lm{3}+\lm{4})(\Hd{1}\H{1})(\Hd{2}\H{2}) - \lm{4}\left| 
\Hd{1}\H{2}\right|^{2} \nonumber\\
  &  & + (\lm{5}\Hd{1}\H{1} +\lm{6}\Hd{2}\H{2})N^{\star}N+
\lm{7}(\H{1}\H{2}N^{\star 2}+h.c.)  \nonumber  \\
 &  &+  \lm{8}(N^{\star}N)^2+
\lambda \mu (N+h.c.)(\Hd{1}\H{1}+\Hd{2}\H{2})  \nonumber \\
 & & + \m{1}\Hd{1}\H{1}+\m{2}\Hd{2}\H{2} + \m{3}N^{\star}N -  m_4 
(\H{1}\H{2}N+h.c.)
 \nonumber\\
 & & - \frac{1}{3}m_5(N^3+h.c.)+\m{6}(\H{1}\H{2}+h.c.)+ 
\m{7}(N^2+h.c.)
\end{eqnarray}
where the quartic couplings $\lambda_i, i =1 \dots 8 $ at the electroweak 
scale are related via renormalization group equations to the gauge couplings 
and the $\lambda, k$ of the superpotential at the supersymmetry 
breaking scale $M_S$, taken to be 1 TeV. 
The boundary values at $M_S$ of the quartic couplings are given by  
$$\lm{1} = \lm{2} = \frac{1}{4}(g_2^2+g_1^2),        
 \lm{3} = \frac{1}{4}(g_2^2-g_1^2), \lm{4} = \lambda^2-\frac{1}{2}g_2^2 ,$$

$$\lm{5}=\lm{6}=\lambda^2,\lm{7}= -{\lambda}k,\lm{8}= k^2. $$

The soft Susy-breaking terms $m_i, i = 1 \dots 7 $, are taken as 
phenomenological 
parameters, without assuming they evolve perturbatively from a more or less 
universal high energy form. The $m_6^2$ and $m_7^2$ terms are absent in 
the theory with $Z_3$ symmetry.

The two Higgs doublets and the singlet are expressed in terms of real fields
\mbox{\p{i}, (i = 1,2,\ldots,10)}, through 

\begin{equation}
\H{1} = \left( \begin{array}{c}H^0_1 \\ H_1^- \end{array} \right) \, , 
\H{2} = \left( \begin{array}{c}H_2^+ \\ H_2^0 \end{array} \right)
\end{equation}

\begin{equation} \label{eq.doublets}
\H{1} =\frac{1}
{\sqrt{2}}\left( \begin{array}{c} \p{1} + i\p{4}\\ 
\p{7} -i\p{9} \end{array} \right) , \; 
\H{2} = \frac{1}{\sqrt{2}}\left(\begin{array}{c} \p{8} + i\p{10}\\ \p{2} + 
i\p{5} \end{array} \right), 
\nonumber\\
N = \frac{1}{\sqrt{2}}\left(  \p{3} + i\p{6}\right) .
\end{equation} 

Taking  real coupling constants, so that the tree level potential is CP 
conserving, but allowing  complex vevs for the  neutral fields,
\be
\langle H_i^0 \rangle = v_i e^{i \theta_i} (i=1,2), \langle N \rangle=v_3 
e^{i \theta_3},
\ee
 gives
\begin{eqnarray} \label{eq.vsvac}
V_{0} &  = &  \frac{1}{2}(\lm{1}v_1^4  +\lm{2}v_2^4) +  (\lm{3}+\lm{4})v_1^2 
v_2^2  +
(\lm{5}v_1^2 +\lm{6}v_2^2)v_3^2  \nonumber\\ 
&  & + 2\lm{7}v_1 v_2 v_3^2 cos(\theta_1 + \theta_2 - 2 \theta_3)+ \lm{8}
v_3^4 +
2 \lambda\mu (v_1^2+v_2^2)v_3 cos(\theta_3)  \nonumber\\
 &  & + \m{1}v_1^2+\m{2}v_2^2 + \m{3} v_3^2-2m_4 v_1 v_2 v_3 cos(\theta_1 
+ \theta_2 + \theta_3) 
 \nonumber\\
 & & - \frac{2}{3} m_5 v_3^3 cos(3 \theta_3)+ 2\m{6}v_1 v_2cos(\theta_1 + 
\theta_2) +2\m{7} v_3^2 cos(2 \theta_3)
\end{eqnarray}
where, without loss of generality,  $\theta_2 = 0$.
As the $m_i$ are unknown we choose five, $m_i$ (i=1,2,3,4,7) to ensure that 
$V_0$ has a stationary value at prescribed magnitudes and phases 
$v_1,v_2,v_3,\theta_1, \theta_3$ using the conditions 
\be \label{eq.mincon1}
\frac{\partial V_0}{\partial v_{i}}=0\,,i=1,2,3  
\ee
\be \label{eq.mincon2} 
\frac{\partial V_0}{\partial \theta_{i}}=0\,,i=1,3  .\\
\ee 

A sixth mass, $m_6$, can be exchanged for the tree level mass of the charged 
Higgs, $M_{H^+}$, which, using eq. (\ref{eq.mincon2}) is 
\begin {eqnarray}
M_{H^+}^2 & = & - \lambda_4 v_0^2 - \frac{2 (\lambda_7 v_3^2 \sin (3 \theta_3) 
+ m_6^2 \sin\theta_3 )}{\sin(2\beta) \sin(\theta_1+\theta_2+\theta_3)}. 
\end{eqnarray}
This shows how the parameter $m_6^2$, not necessarily positive, and absent 
in the $Z_3$ symmetric case, introduces extra freedom to raise the charged 
Higgs mass.
This leaves one parameter $m_5$, with no particular interpretation.  Fixing 
$v_0 = \sqrt{v_1^2+v_2^2} = 174 \mbox{ GeV}$, we take as parameters 
$\tan \beta \equiv v_2/v_1$, \mbox{ $R \equiv v_3/v_0 $}, 
$\theta_1, \theta_3, M_{H^+}$ and $m_5$. There are also $\lambda, k$ and 
$\mu$ in the superpotential.

Sets of parameters are chosen which satisfy eqns.(\ref{eq.mincon1},
\ref{eq.mincon2}), and the 
mass matrix is calculated. Cases with positive mass squared are accepted, as 
these correspond to a local minimum. 
The scalar mass-squared matrix is given by
\be \label{eq.massmat} M_{ij}^{2}=\frac{\partial V_0}{\partial \phi_{i}
\partial \phi_{j}}\,\left|_{\phi = \langle\phi\rangle}\right. \,,
(i,j=1,10).  \ee
The 6x6 neutral block describes 1 zero mass would-be Goldstone boson and 5 
massive physical particles, of which 3 are CP even and 2 CP odd when CP is 
conserved.

\section{Higgs Spectrum} \label{MHiggs}

\begin{figure}[thb]
\unitlength1cm
\begin{minipage}[t]{5.5cm}
\begin{picture}(5.5,5.5)
\epsfig{file=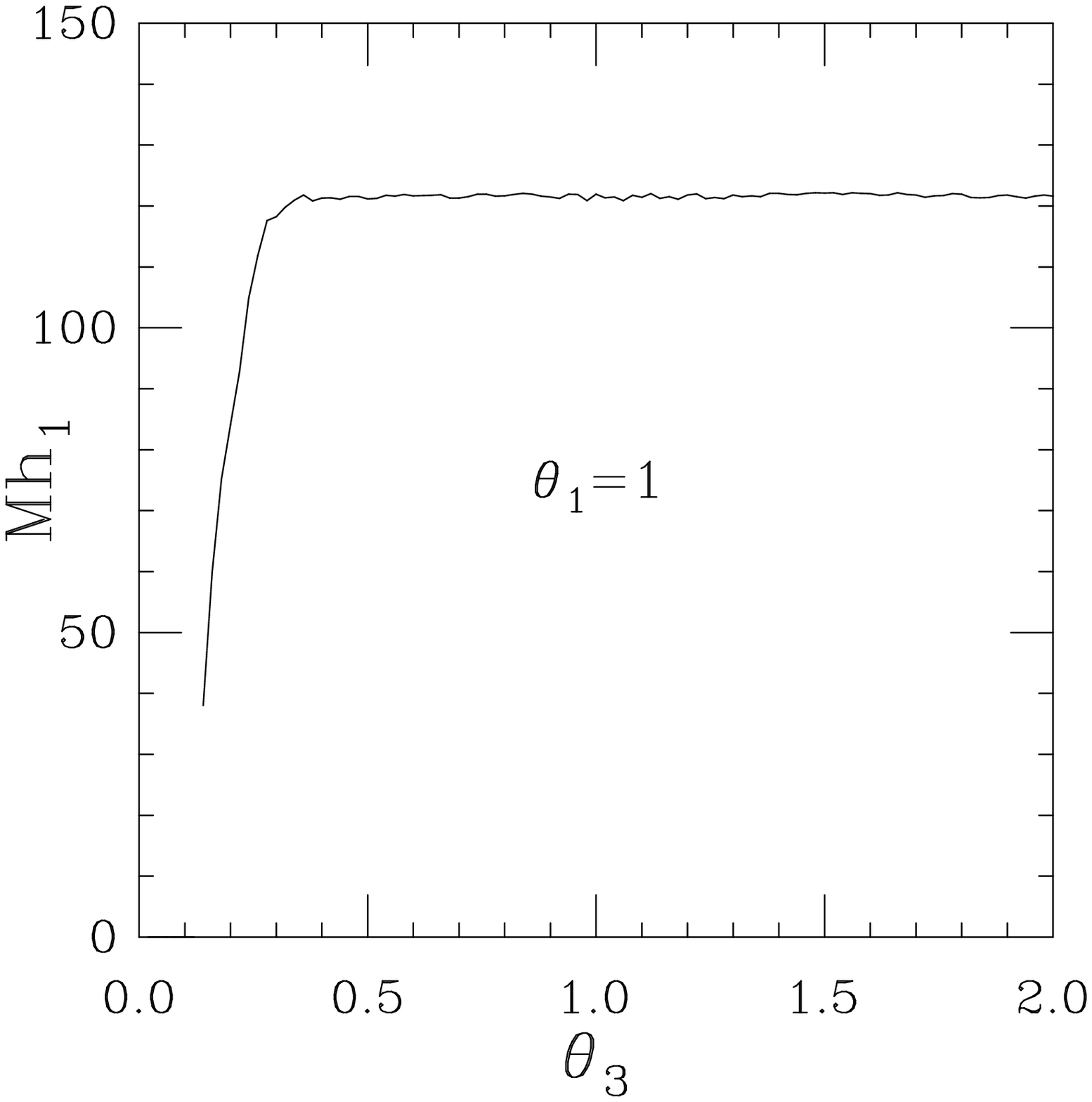, width=5.5cm}
\end{picture}\par
\end{minipage}
\hfill 
\begin{minipage}[t]{5.5cm}
\begin{picture}(5.5,5.5)
\epsfig{file=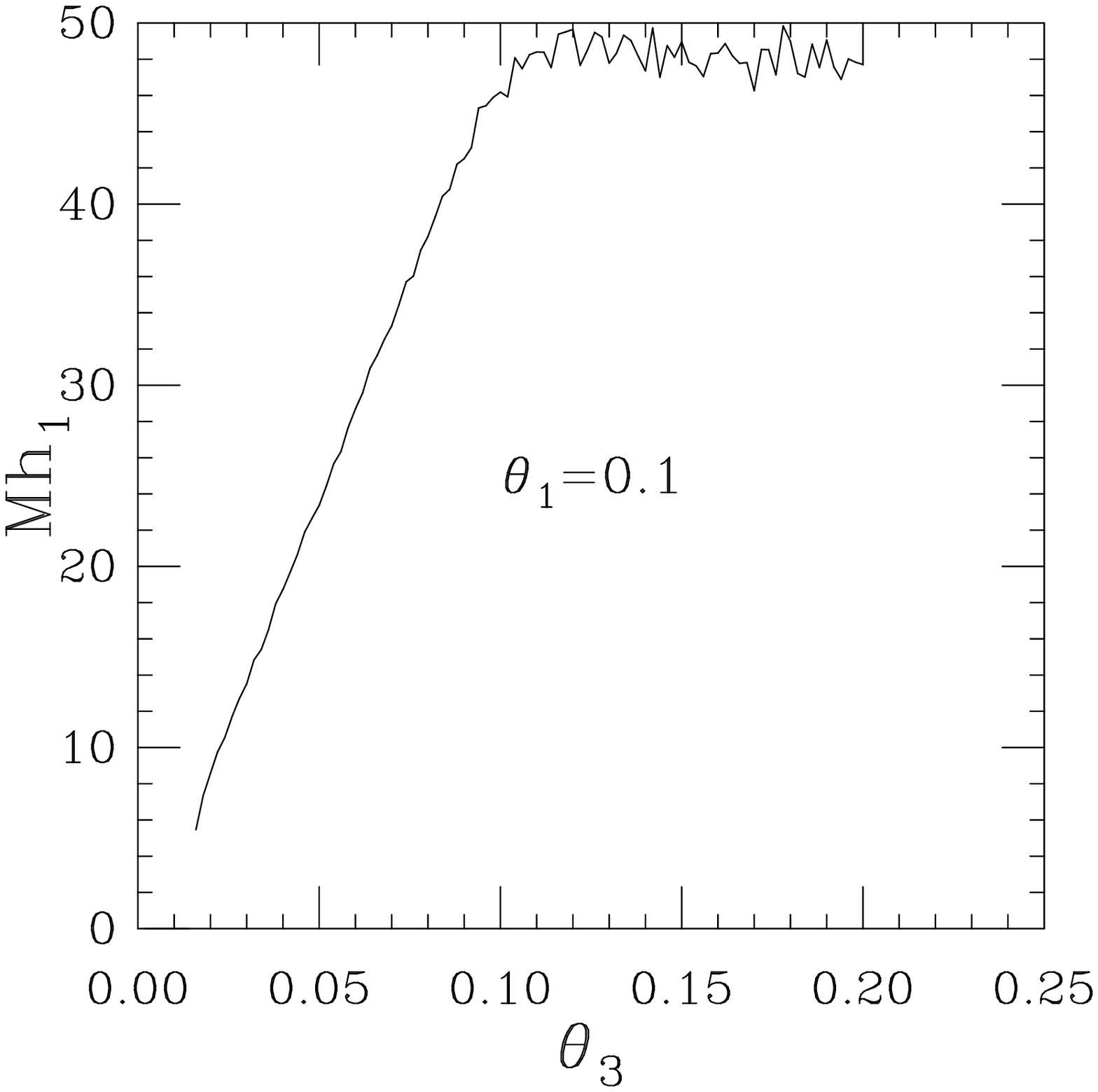, width=5.5cm}
\end{picture}\par
\end{minipage}


\unitlength1cm
\begin{minipage}[t]{5.5cm}
\begin{picture}(5.5,5.5)
\epsfig{file=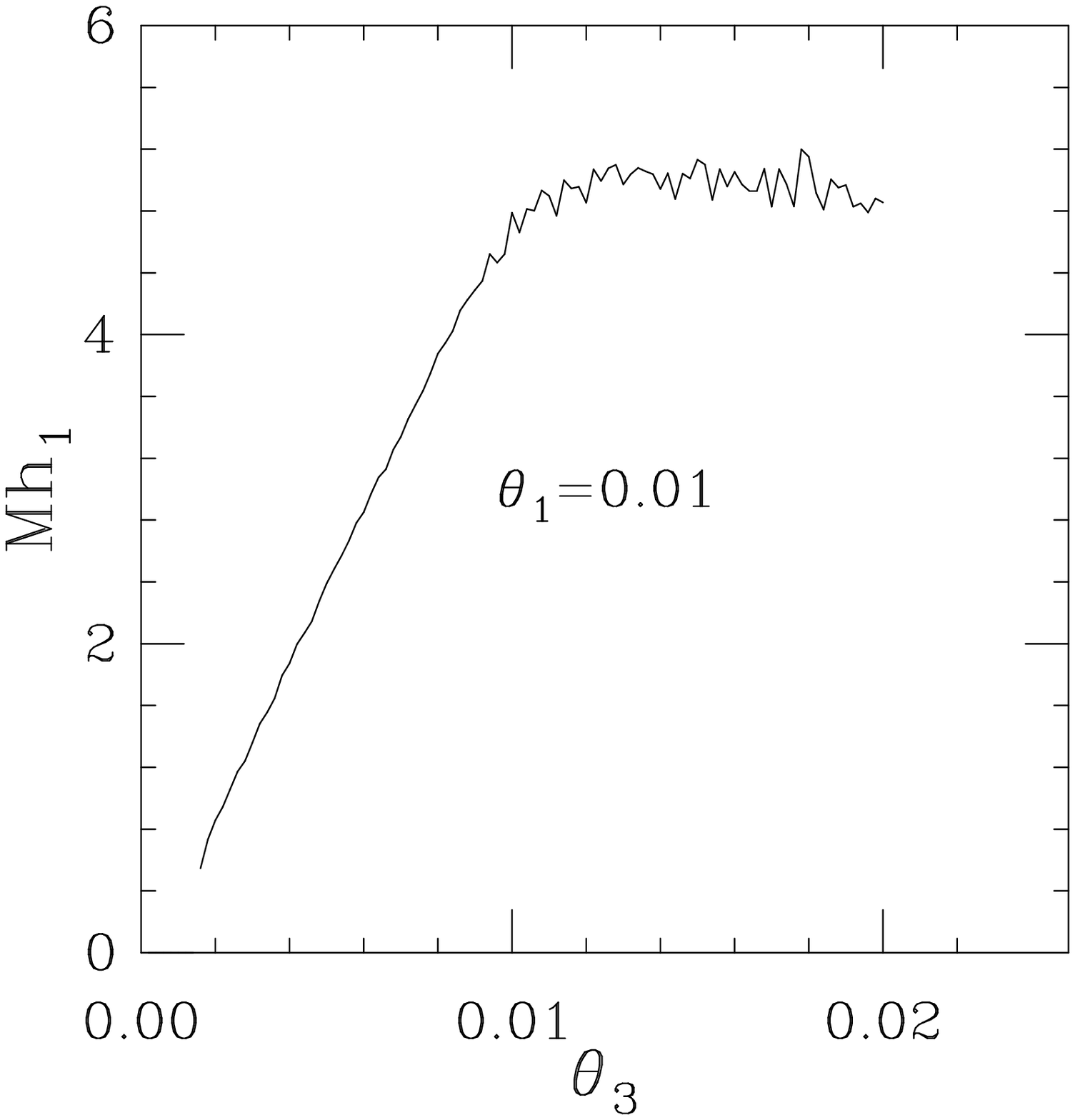, width=5.5cm}
\end{picture}\par
\end{minipage}
\hfill 
\begin{minipage}[t]{5.5cm}
\begin{picture}(5.5,5.5)
\epsfig{file=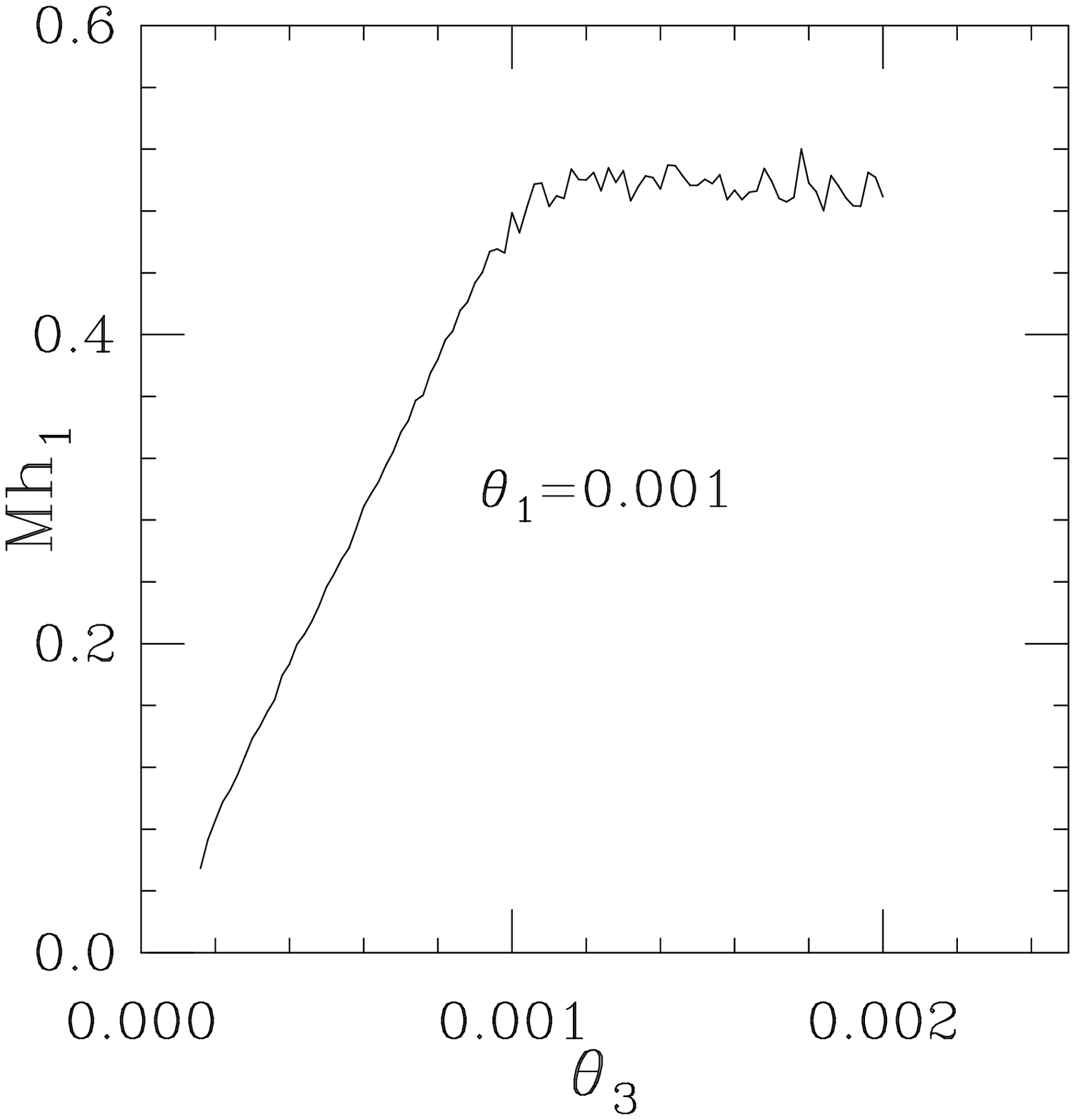, width=5.5cm}
\end{picture}\par
\end{minipage}

\caption{Upper bound on the lightest neutral Higgs mass $M_{h_1}$ 
as a function of $\theta_3$, for 
$\theta_{1}$ equal to 1, 0.1, 0.01 and 0.001 radians respectively and with 
Susy breaking scale $M_S$ = 1 TeV.
\label{Fig: massbounds}}
\end{figure}

It has been shown that at tree level the NMSSM 
with $Z_3$ symmetry and with arbitrary soft terms 
does not allow SCPV~\cite{Romao}. However, the inclusion 
of radiative corrections does allow SCPV together with a light boson. 
Various treatments 
of the radiative corrections and Susy-breaking 
potential~\cite{Babu,Haba,Asatrian} produce different 
Higgs spectra, some of which can now be excluded 
by experiment, but less readily than  in the MSSM, 
due to dilution of couplings by the singlet field. 
As noticed by Pomarol~\cite{Pom2} inclusion of general 
$Z_3$ symmetry breaking terms does allow SCPV.  We have 
investigated the mass spectrum for the potential of 
eq.(\ref{eq.vs0}) including such terms and found that it is 
quite possible to produce an experimental spectrum with no light 
particles~\cite{DavJer}, but all these solutions had large 
CP violating phases. 
Such phases in the NMSSM, as in the MSSM, give rise to large contributions to 
electric dipole moments. They can be suppressed if the squark masses are 
several TeV, or if contributions from different diagrams cancel. We have 
calculated the neutron and electron EDMs choosing the unknown soft parameters 
at random, and found that very few sets produced the necessary cancellations. 
We therefore explored further the possibility that the phases 
are small. In analyses where SCPV is induced by radiative corrections 
small phases arise naturally and are accompanied by a light scalar, 
as expected by the Georgi-Pais theorem. In our case SCPV occurs at tree
 level, so this theorem does not apply. Nevertheless, when we require 
the phases to be small, and use these as input to fix some parameters 
in the potential, we find that that there is always a light 
Higgs particle $h_1$. 
Figure \ref{Fig: massbounds} shows the upper bound on the 
lightest neutral Higgs mass $M_{h_1}$. 
Each graph is for a set of values of $\theta_3$ increasing from 0 to 
$2\theta_1$ where $\theta_1$ is fixed at 1, 0.1, 0.01 and 0.001 radians. 
For each 
$\theta_1,\theta_3$ we randomly selected 
100,000 sets of the other parameters with values in the following ranges: 
$2\le \tan\beta \le 3$, $10 \le v_{3}\le 510$ GeV, 
$-500\le m_{5} \le 500$ GeV, 
$-500\le \mu \le 500$ GeV, $55 \le M_{H^{\pm}}\le 800$ GeV 
and $\lambda = k =0.5$.

These figures and those below correspond to local minima, 
but experience suggests that,
 given enough computer time, global minima could be found giving 
masses not far below these bounds.
Such graphs show that the upper bound when 
$\theta_1$ and $\theta_3$ are small ($\lsim 0.1$ radians) is roughly 
$M_{h_1}\simeq \frac{\min (\theta_1, \theta_3)}{0.01}\, 5$ GeV.

\section{Light Higgs} \label{GPThm}

The result that weak  spontaneous CP breaking implies a light Higgs particle 
is quite general, and may be understood by a variant on the 
argument of Georgi and Pais~\cite{Georgi}.

The central step in the proof  of the Georgi-Pais theorem is the equation

\be \label{eq.GP}
\sum_{k}\,\frac{\partial^2 V_o}{\partial\phi_{j}\partial\phi_{k}}
\,\left( (U\delta\lambda )_{k}-\delta\lambda_{k}\right) =0, 
\ee
where $V_o(\phi )$ is the field dependent scalar potential, $\phi_{j}$ are 
spinless boson fields, the vector $\lambda$ is the value of $\phi$ 
at which the minimum of 
the scalar potential occurs, U is the CP symmetry operator and 
$\delta\lambda$ 
is the change due to radiative corrections.  If 
there is no spontaneous symmetry breaking, $U\lambda =\lambda$ and 
from eq.(\ref{eq.GP}) it is clear that, if the mass matrix 
$\partial_j \partial_k V_0$ is 
not singular, the relation 
$U\delta\lambda =\delta\lambda$ holds and 
no SCPV occurs. On the other hand if a massless particle is present 
in the unbroken theory then 
the mass matrix is singular and SCPV may be induced by the 
radiative corrections. The massless mode will gain a small 
mass as a result of the radiative corrections. It is this 
mechanism which produces the light particle when SCPV is induced in the 
MSSM or the NMSSM with the $Z_{3}$ discrete symmetry.

The key assumption is that the breaking is perturbative. 
We make a corresponding assumption, in the general NMSSM model 
without $Z_3$ symmetry, 
that there is a SCPV minimum with small phases and that the 
effective potential can be expanded as a Taylor 
series about this point. For small CP violating phases the 
potential of eq.(\ref{eq.vs0}) has CP conjugate 
minima at the points

\begin{equation}
\underline{\epsilon}_1 =  (v_1,v_2,v_3, v_1 \theta_1, v_2 \theta_2, v_3 
\theta_3), 
\end{equation}
\begin{equation}
\underline{\epsilon}_2= (v_1,v_2,v_3,-v_1 \theta_1,-v_2 \theta_2,-v_3 
\theta_3)
\end{equation} 
in a basis of $(Re H_1^0,Re H_2^0,Re N,ImH_1^0,Im H_2^0, Im N)$. 
Performing a  Taylor expansion of $\frac{\partial V}{\partial \phi_{i}}$ 
about $\underline{\phi} = \underline{\epsilon}_{1}$
\begin{eqnarray}
(\epsilon_2 - \epsilon_1)_j \frac{\partial^2 V}{\partial \phi_j \partial 
\phi_i}\Big|_{\underline{\epsilon}_1} \approx
\frac{\partial V}{\partial \phi_i} \Big|_{\underline{\epsilon}_2} - 
\frac{\partial V}{\partial \phi_i} \Big|_{\underline{\epsilon}_1}
= 0 - 0,
\end{eqnarray} 
we see that to leading order the mass squared matrix must be singular. 
To this order 
there is a zero mass particle, with eigenvector along the direction 
$\underline{\epsilon}_2-\underline{\epsilon}_1$ in the 
6-dimensional neutral Higgs space joining the two CP violating minima. If 
$\theta_i \neq 0$, the neutral matrix does not decouple into sectors with 
CP = +1 and -1, but it does so approximately, as the off diagonal blocks 
of the matrix are proportional to the small angles $\theta_i$. To this 
approximation, the light particle is always in the matrix block corresponding 
to imaginary parts of the fields and so is almost purely CP odd. 
Depending on the 
parameters in the potential, this particle can be a varying admixture of 
singlet and doublet fields.

This is most easily examined by changing to the unitary gauge. In the 
limit of zero phases, the matrix reduces to a 3x3 block of the real parts 
of the fields and a 2x2 block of the imaginary parts of fields of the form
\be \left( \begin{array}{lll} \frac{A}{\sin\beta \cos\beta} &  &  
\frac{v_{0}}{v_{3}}B \\   &  &  \\ 
\frac{v_{0}}{v_{3}}B & & C \end{array} \right) \label{massmatr}  \ee \\ 
where A is the (5,4) element of the whole 6x6 mass matrix, 
B is equal to $\frac{v_{2}}{v_{3}}$ times the (6,4) element, 
and C is equal to the (6,6) element. In the matrix of eq.(\ref{massmatr}) 
the (1,1) element is doublet and the (2,2) element is singlet. 

The condition for a massless pseudoscalar is obviously
\be
 \frac{AC}{\sin \beta \cos \beta}  = (\frac{v_0}{v_3}B)^2. 
\label{ACB}
\ee
When the phases are not zero but small, this particle becomes 
light and almost-pseudoscalar. 
We can readily see  how the nature of the light particle depends on 
$\theta$. For example, if $\theta_1<< \theta_3$ the eigenvector 
$\underline{\epsilon}_2-\underline{\epsilon}_1$ is in the singlet 
direction, and the 
massless particle is pure singlet.  
In this case eq.(\ref{eq.mincon2}) 
just gives B = A = 0 , and so eq.(\ref{ACB}) is satisfied with $C\ne 0$. 
Likewise for $\theta_3<< \theta_1$ 
eq.(\ref{eq.mincon2}) gives B = C = 0 and the light particle is pure 
doublet. 
In general the N field percentage in the light pseudoscalar eigenvector 
is 
\be \label{Nformula} N\%\simeq\frac{100\,v_{3}^{2}\,\theta_{3}^{2}}
{v_{0}^{2}\,\sin^{2}\beta\,\cos^{2}\beta\,
\theta_{1}^{2}+v_{3}^{2}\,\theta_{3}^{2}} \ee
which is small if $v_3\theta_3 << v_0 \theta_1$.
For high values of $v_{3}$ such that 
$v_{0}$ can be neglected, the N field fraction
will be independent of $\theta_{3}$ and $v_{3}$ and equal to 1. It is 
significant for the experimental detectability that the fraction of 
singlet is naturally high, even for quite moderate values of the parameters.
For example, tan$\beta$ = 1, $v_{3}=v_{0}$, $\theta_{1}=\theta_{3}$, 
gives N\% = 80\%, and tan$\beta$=2, $v_{3}=2\,v_{0}$, 
$\theta_{1}=\theta_{3}$ gives a light pseudoscalar which is 96\% singlet. 

The result that small phases imply a light particle is not specific to the 
NMSSM. It is more transparent in a general 2 doublet model 
with no explicit CP violation,
 with doublets $\Phi_1$ and $\Phi_2$, where we need 
keep only one SCPV phase $\theta$ and  
analytical formulae can be given for all tree level masses. In such a model 
with a discrete $Z_2$ symmetry $\Phi_1 \rightarrow -\Phi_1, \Phi_2 
\rightarrow \Phi_2$, the equation for a stationary value of the 
effective potential is
\be 
\sin\theta (2 \lambda_5 v_1 v_2 \cos \theta - m_{12}^2)=0,
\ee
where $m_{12}^2 \Phi_1^\dagger \Phi_2$ and  
$\lambda_5 (\Phi_1^\dagger \Phi_2)^2$ are terms in the potential. 
So as well as the CP conserving solution
 \be \label{CPC} \sin \theta =0\,, \ee
there is the SCPV one at $\pm \theta$, where
 \be \label{eq.SCPV}
\cos \theta= \frac{m_{12}^2}{2 \lambda_5 v_1 v_2}.
 \ee
 The mass of the pseudoscalar A when $\theta =0$ is
 \be 
M_A^2 = v_0^2 (\frac{m_{12}^2}{2 v_1v_2} -\lambda_5),
\ee 
and is a continuous function of $\theta$ in the SCPV case. 
So we see from eq.(\ref{eq.SCPV}) 
that small $\theta$ implies a low mass $M_A$. On the other hand, the CP 
conserving eq.(\ref{CPC}) does not imply a zero mass particle. In the 
MSSM $\lambda_5 = 0$ at tree level and $m_{12}^2$ is a free parameter, which
can be large. Radiative corrections generate a small 
$\lambda_5$~\cite{Maekawa}, and SCPV becomes possible, but, 
as shown by eq.(\ref{eq.SCPV}), only if $\frac{m_{12}^2}{2 v_1v_2}$ 
is also small.

Here the equations are identical to the classical dynamics problem of a 
bead free to slide on a vertical circular wire, radius $a$, constrained 
to rotate about its vertical diameter at constant angular velocity $\omega$. 
For small $\omega$ the stable equilibrium position is $\theta=0$ but if 
$\omega$ exceeds a critical value $\sqrt{\frac{g}{a}}$ the stable 
equilibrium is at $\theta \ne 0$. The angular frequency of oscillation 
about this position is $p =\omega$ sin $\theta$, which tends to zero  
for small $\theta$. Oscillations about the stable $\theta = 0$ position 
have $p = \sqrt(\frac{g}{a}-\omega^2)$, which tends to zero at the 
critical $\omega$ but can be large near $\omega = 0$.

These examples clarify how in the NMSSM case small $\theta$ implies a light 
particle, but $\theta  = 0$ need not. They also bring out the point implicit 
in our approximate argument that there are no large parameters 
in the problem. In the dynamics case there would be no low frequency 
mode if $\omega$ were O$(\theta^{-1})$.  In the NMSSM the argument breaks 
down if the vev ratio $v_3/v_{0}$ is very large.

\section{N field and second lightest neutral Higgs boson for small 
phases} \label{Numerical}

In the results presented below and in Fig.\ref{Fig: massbounds},
we have diagonalized the full 6x6 neutral mass 
squared matrix numerically. We have then made an 
orthogonal transformation to isolate the 
Goldstone mode, in order to look at the 
eigenvectors of the 5 physical particles $h_i$, to 
determine their field content and to obtain the ZZ$h_i$ and Z$h_ih_j$ 
couplings.

\begin{figure}[thb]
\unitlength1cm
\begin{minipage}[t]{5.5cm}
\begin{picture}(5.5,5.5)
\epsfig{file=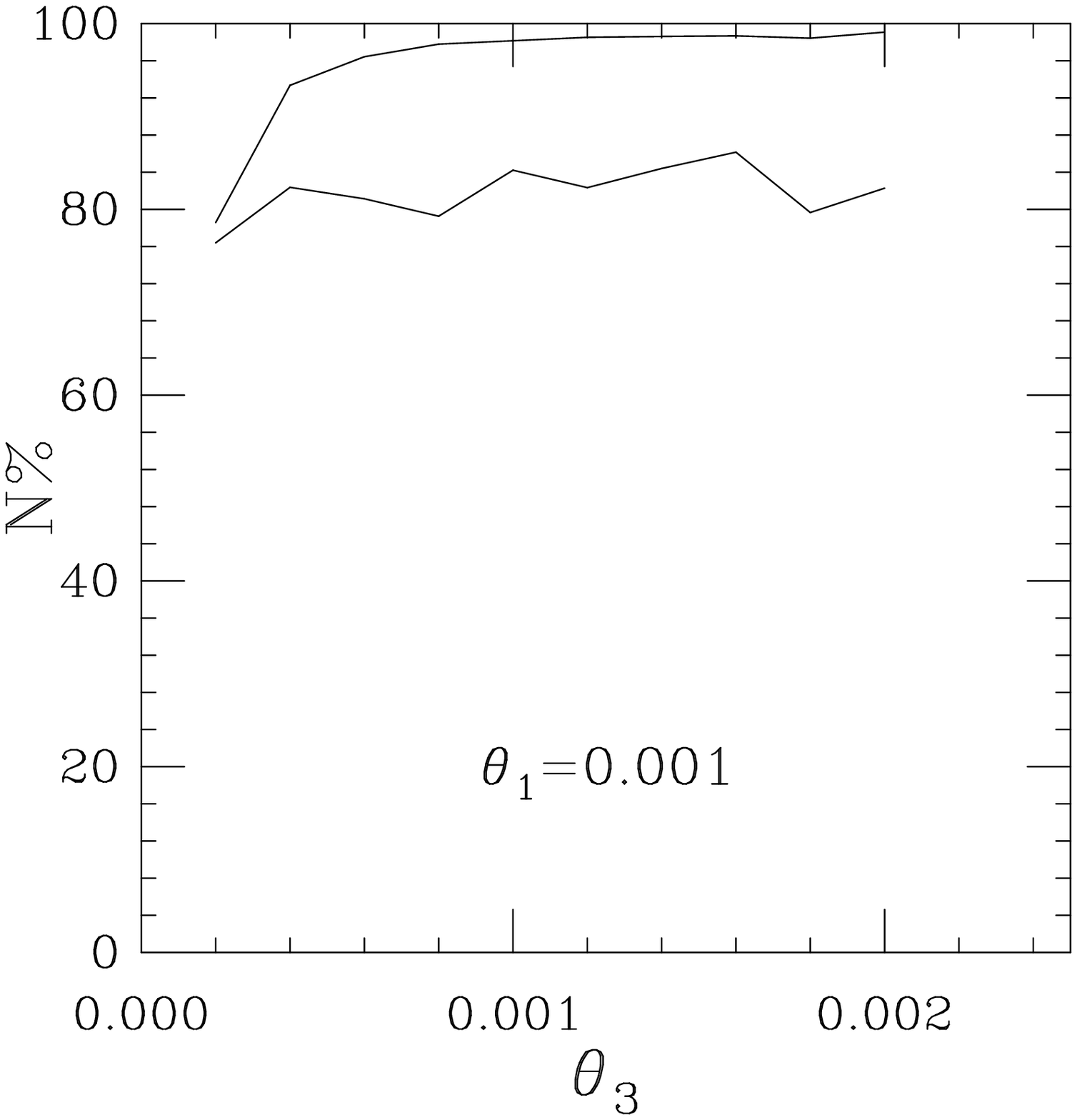, width=5.5cm}
\end{picture}\par
\caption{N field percentage in the eigenvector of the 
lightest neutral Higgs boson 
for $M_{H^{\pm}}$=200-800 GeV (upper line) and $M_{H^{\pm}}$=55-200 GeV 
(lower line) for $\theta_{1}$=0.001, $\theta_{3}=0-2\theta_{1}$.
\label{NFig1}}
\end{minipage}
\hfill 
\begin{minipage}[t]{5.5cm}
\begin{picture}(5.5,5.5)
\epsfig{file=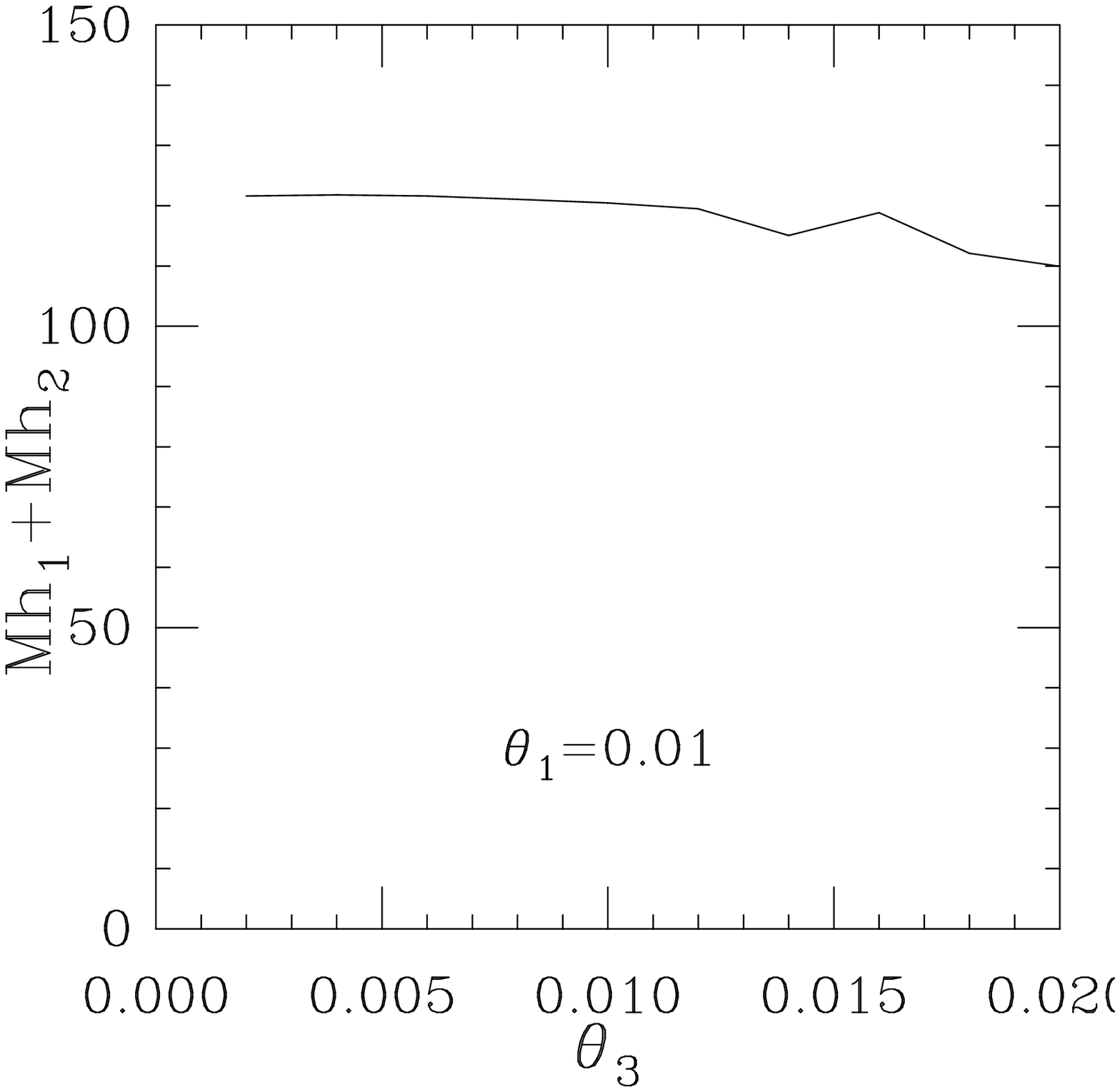, width=5.5cm}
\end{picture}\par
\caption{The sum of the masses of the two lightest neutral Higgs 
bosons $M_{h_1}$+$M_{h_{2}}$ as a function of $\theta_{3}$, with
$\theta_{1}$=0.01 radians, $M_{H^{\pm}}$=55-200 GeV and 
$M_S$=1 TeV.
\label{mhma2}}
\end{minipage}
\end{figure}

The first important result is that described in Section \ref{GPThm}: the 
existence of  a light particle $h_1$ when the phases are 
small enough to naturally generate electron and neutron dipole 
moments consistent with experiment.
Its eigenvector 
is almost entirely  composed of the imaginary parts of the doublet 
$H_1$, $H_2$  and singlet N fields. 
The percentage of singlet in the eigenvector of $h_1$ is high in general, 
as can be seen in Fig. \ref{NFig1}. 
The exact N field percentage depends on the value of $v_3/v_{0}$.  
This is correlated 
with the value of $M_{H^+}$. Although these parameters  are independently 
specified, it is found that the condition that all masses are real 
forces $M_{H^+}$  and $v_3$ to increase together. 
This favours a high singlet content, which is crucial as far as 
possible experimental detection of the pseudoscalar $h_1$ is concerned.

The second lightest neutral Higgs boson $h_2$ is always nearly pure scalar 
and is mainly doublet in the 
region of the parameter space where $v_3$ and $M_{H^{\pm}}$ 
are large compared to $v_{0}$: it could therefore be visible in 
future experiments. Furthermore, 
the sum of the masses of the lightest and second lightest neutral Higgs 
bosons has an upper bound of about 120 GeV as shown in Fig.\ref{mhma2}. 
In these graphs radiative corrections have been incorporated 
using the Standard Model renormalisation group equations 
between $M_S$ = 1 TeV and the electroweak scale $v_0$ = 174 GeV, for 
$\tan \beta$ between 2 and 3. These radiative corrections 
correspond to having degenerate stops with a mass of $M_S$ = 1 TeV 
and are significant for the nearly pure scalar mass $M_{h_2}$.

\section{Experimental Signature} \label{Expt} 

One might think that a light boson such as we predict has been ruled out 
by experiment, but this is not so. 
Our light particle is difficult to detect for two reasons.
In the first place it is almost purely CP odd, and there is 
no ZZA coupling, where A 
denotes a CP=-1 particle; S will be used for one with CP=+1. A ZA 
state can not be 
produced at LEP, for real or virtual Z, 
but associated production of SA is possible if kinematically allowed. 
This was not seen at LEP, possibly because the 
ZSA coupling was small. In the MSSM the ZZS coupling is 
complementary to the ZSA coupling 
so that non production of ZS allows LEP2 to exclude 
$M_A \le 80$ GeV, $M_S\le 80$ GeV 
~\cite{Gay}. Within the MSSM, therefore, experiment excludes a 
light pseudoscalar. This complementarity argument 
does not apply to a general 2 doublet model or to the NMSSM.
The second obstacle to detection is that the light near-pseudoscalar 
can be mainly 
singlet, and a singlet does not couple to gauge fields, quarks or 
leptons at tree 
level, only to the other Higgs particles.  
The singlet component will not contribute to $\Upsilon$ decays 
by the Wilczek process, and thus with N$\%\gsim$ 90$\%$ and tan$\beta\lsim$ 3 
the lower limits \cite{KRA} on the mass of a pseudoscalar can be evaded. 
To get a more quantitative idea of the experimental detectability, we have 
calculated the production cross section of all the Higgs bosons at LEP2 
energies. Each set of parameters gives definite tree level masses and 
couplings for all the Higgs particles. We have not attempted 
a Monte Carlo analysis, 
in view of the uncertainties and  complexities of decay modes and detector 
efficiencies. Following the spirit of one of the few analyses 
based on the NMSSM ~\cite{King}, we arbitrarily assume that 
if the cross section 
in any production channel $Zh_i$ or $h_ih_j$ was as large as 0.3 pb, 
corresponding to 20 events at LEP2 luminosity of 
175$\times$4 pb$^{-1}$ 
at 200 GeV, a signal would have been detected. Although for $\theta_{1}$ and 
$\theta_{3}< 0.1$ all our parameter sets gave a neutral Higgs boson of 
mass $<$50 GeV, we could always find regions of the 
($M_{H^{\pm}}$, $v_{3}$, $m_{5}$, $\mu$, $\tan\beta$) 
 space in which this particle and its partners are undetectable at LEP, 
except for $\theta_{3}<<\theta_{1}$ where the particle is mainly doublet.

\section{Conclusions}
Spontaneous CP violation is possible in the NMSSM at tree level. 
It can give an acceptable 
Higgs spectrum. If the phases on the vevs are small, 
there is a light particle, predominantly 
pseudoscalar, and predominantly singlet in much of the parameter space, and
thus hard to detect. Phases  $\theta_{1,3} = {\cal{O}}(0.01)$, 
such as may be required to 
suppress electric dipole moments, give a mass 
$M_{h_1} \lsim \frac{\min(\theta_1, \theta_3)}{0.01}\, 5$ GeV. 
This model is not 
yet ruled out but will be open to more stringent tests at higher energy
 colliders such as the LHC, where all 5 neutral Higgs bosons 
should be kinematically accessible and their couplings cannot all be 
predominantly singlet.

\section*{Acknowledgements}
We would like to thank D.G. Sutherland for discussions. 
This work was supported in part by PPARC and the EU TMR Network 
contract FMRX-CT97-0122.

%


\begin{thebibliography}{99}
\bibitem{Krauss} L.M. Krauss and S.J. Rey, \Journal{\PRL}{69}{1308}{1992}
\bibitem {IbNath} T. Ibrahim and P. Nath, \Journal{\PRD}{57}{478}{1998}.
\bibitem{Kane} M. Brhlik, G J. Good and G.L. Kane,
 \Journal{\PRD}{58}{115004}{1999}. 
\bibitem{Abel} S. Abel, S.Khalil, and O. Lebedev, hep-ph/0103320.
\bibitem{Maekawa} N. Maekawa, \Journal{\PLB}{282}{392}{1992}.
\bibitem{Pom1} A. Pomarol, \Journal{\PLB}{287}{331}{1992}. 
\bibitem{Georgi} H. Georgi and A. Pais, \Journal{\PRD}{10}{1246}{1974}.
\bibitem{Gay} P. Gay, {\it~Proc. of Int. Europhysics Conference on High 
Energy Physics}, Tampere, 1999, Eds. K. Huito, 
H. Kurki-Suonio, J. Maalampi, IoP. 
\bibitem{Romao} J.C. Rom\~ao, \Journal {\PLB}{173}{309}{1986}.
\bibitem{Babu} K.S. Babu and S.M. Barr, \Journal{\PRD}{49}{R2156}{1994}.
\bibitem{DavJer} A.T. Davies, C.D. Froggatt and A. Usai,{\it~ Proc. of the 
International Europhysics Conference on High Energy Physics}, Jerusalem 
(1997), Eds. D. Lellouch, G. Mikenberg and E. Rabinovici , Springer-Verlag 
(1998),p891. 
\bibitem{DavCop} A.T.Davies, C.D. Froggatt and A. Usai, 
{\it~ Proc. of Conference on Strong and Electroweak 
Matter, Copenhagen} (1998), Eds. J. Ambjorn, P.H. Damgaard, 
K. Kainulainen, K. Rummukainen, World Scientific (1999), p227.
\bibitem{Branco} G.C. Branco, F. Kr\~uger, J.C. Rom\~ao, A. Teixeira, 
hep-ph/0012318.
\bibitem{Gunion} J.F. Gunion, H.E. Haber, 
\Journal{\NPB}{272}{1}{1986}.
\bibitem{Higgshunter} J.F. Gunion, H.E. Haber, G. Kane, S. 
Dawson, The Higgs Hunter's Guide, (Addison-Wesley, Reading MA, 1990).
\bibitem{Haba} N. Haba, M. Matsuda and M. Tanimoto,~\Journal{\PRD}{54}{6928}
{1996}.
\bibitem{Asatrian}  H.M. Asatrian, G.K. Eguian, 
Mod. Phys. Lett. A10, (1995), 2943.
\bibitem{Pom2} A. Pomarol, \Journal{\PRD}{47}{273}{1993}.
\bibitem{KRA} M. Krawczyk, hep-ph/0103223.
\bibitem{King} S. F. King and P. L. White, \Journal{\PRD}{53}{4049}{1996}.




\end{thebibliography}
\end{document}